\documentclass[11pt]{article} \usepackage{amsfonts}   

\makeatletter
\def\quality{\textheight=240mm \textwidth=160mm \topmargin=0Truein
             \ifcase \@ptsize \hoffset=-23mm
                     \or \hoffset=-20mm \or \hoffset=-15mm \fi}
\def\bdraft{\pagestyle{myheadings} 
           \textheight=10.5truein \textwidth=7.5truein \parindent=8pt
           \voffset=-1truein \topmargin=30pt \headheight=10pt \headsep=3pt
           \ifcase \@ptsize \hoffset=-1.5truein \or \hoffset=-1.35truein
                        \or \hoffset=-1.15truein \fi}
\makeatother
\quality

 \def\n{\noindent} 
 \def\IZ{{\mathbb{Z}}}
  
   \def\cA{{\cal A}}
\def\toas#1{\stackrel{#1}{\longrightarrow}}

\def\intp#1{\left\lfloor#1\right\rfloor}  \def\intpm#1{\lceil#1\rceil}
              \def\CR{$$ $$}
  \def\cC{{\cal C}} \def\X{\mathbb{X}}
 \def\den{\rho}  \def\comment#1{}

\def\bline(#1,#2)(#3,#4)(#5){\put(#1,#2){\line(#3,#4){#5}}}  

\newcommand\mlbscale{1pt} 
\newif\iffigs\figstrue 

\def\bfig(#1,#2)#3#4{\begin{figure} \begin{center}
    \framebox{\setlength{\unitlength}{\mlbscale}
       \iffigs \begin{picture}(#1,#2) #3 \end{picture}
       \else \begin{picture}(60,10)(0,0)
                   \put(0,0){\framebox(60,10){Figure}} \end{picture} \fi}
    \end{center} \caption{#4} \end{figure}}

\def\Bfig(#1,#2)#3#4{\begin{figure} \begin{center}
    \setlength{\unitlength}{\mlbscale}
       \iffigs \begin{picture}(#1,#2) #3 \end{picture}
       \else \begin{picture}(60,10)(0,0)
                   \put(0,0){\framebox(60,10){Figure}} \end{picture} \fi
    \end{center} \caption{#4} \end{figure}}

\def\n{\noindent} \def\IZ{{\mathbb{Z}}}

\def\beq#1#2{\begin{equation} \label{#1} #2 \end{equation}}

\def\thname{Theorem}     \def\lmname{Lemma}      \def\prname{Proposition}
\def\dfname{Definition}  \def\crname{Corollary}  \def\rmname{Remark}

\newtheorem{lemma}{\lmname}[section]     

\newtheorem{dftn}{\dfname}[section]
\newtheorem{rmrk}[lemma]{\rmname}

             \catcode`@=11 \@addtoreset{equation}{section} \catcode`@=12

\begin{document}
\title{On exclusion type inhomogeneous interacting particle systems}
\author{\large Michael Blank\thanks{
               Russian Academy of Sci., Inst. for
               Information Transm. Problems,
               and Observatoire de la Cote d'Azur, ~
               e-mail: blank@iitp.ru}}

\date{August 11, 2006}
\maketitle

\begin{abstract}
For a large class of {\em inhomogeneous} interacting particle
systems (IPS) on a lattice we develop a rigorous method for
mapping them onto {\em homogeneous} IPS. Our novel approach
provides a {\em direct} way of obtaining the statistical
properties of such inhomogeneous systems by studying the far
simpler homogeneous systems. In the cases when the latter can be
solved exactly our method yields an exact solution for the
statistical properties of an inhomogeneous IPS. This approach is
illustrated by studies of three of IPS, namely those with
particles of different sizes, or with varying (between particles)
maximal velocities, or accelerations.
\end{abstract}

\n Key words: interacting particle system, exclusion process,
parallel dynamics, traffic flow model.

\section{Introduction}
A very substantial progress on the understanding of statistical
properties of lattice interacting particle systems (IPS) (see an
excellent review in \cite{Lig} and numerous references therein)
has been achieved mainly for continuous time systems describing
interactions of identical particles. Only recently results related
to non homogeneous particle systems started to show up. Some of
them analyze the presence of a single inhomogeneity (like a street
light in \cite{JL}) or spatially varying hopping rates (see, e.g.
\cite{Ev, GKS}), while in some other papers the situation when a
single particle occupies several lattice sites were considered
(see, e.g. \cite{LC}). In each of these papers the authors
developed new (and quite complicated) constructions or
approximations to deal with the inhomogeneities. Our strategy is
based on a completely different idea, namely we reduce the
analysis of an inhomogeneous problem to a homogeneous one for
which a solution is much simpler or even already known. Returning
to the original setting one is able to recover the complete
statistical description for the inhomogeneous system. In
distinction to complicated mean field approximations the exact
constructions that we use are surprisingly simple and
straightforward. We study three new situations when the analysis
of an inhomogeneous particle system can be reduced to a
homogeneous one. The first of them is the case when the particles
differ in size, i.e. each particle occupies several lattice sites
and, what is more important, the sizes of different particles
might differ. It is worth note that the ability to deal with
particles of different sizes is very important from the point of
view of applications to dynamics of traffic flows (see a review
with numerous references in \cite{CSS}), where ordinary vehicles
and buses or tracks are clearly of different sizes, or to various
biological models like ion channels and mRNA translation where
ribosomes and large molecules or vesicles might be of very
different lengths. The only known result related to such systems
\cite{LC} describing the motion of identical `long' particles is
based on a mean field approximation. Note also that the difference
in particle sizes represents a fundamental obstacle for the
application of one of the basic tools of the IPS theory -- the
coupling method.

As we shall show in Section~\ref{s:size} due to a reduction to a
homogeneous system with particles of unit size the exact solution
is readily available. The main result of this Section is that if
the only difference between the particles in the system is their
size then there exists a bijection (one-to-one correspondence)
between the original system and the system with identical
particles having the same other properties. The idea here to
`compress' `long' particles into `short' ones cutting the `odd'
lattice sites. It might be surprising that details of the dynamics
are not important and the system might be both probabilistic or
deterministic. However, our construction holds only in the case of
the one-dimensional lattice and an exclusion type constraint: not
more than one particle may be present at a lattice site. At
present it is not clear if there are any possible generalizations
to the multidimensional case.

In Section~\ref{s:max-vel} we consider a version of
the model introduced in Section~\ref{s:size} in which in
distinction to the previous setting we assume that each
particle has its own maximal velocity but all of them have
the same unit size. In particular this setting may
appear as a result of the reduction of particle sizes by a
procedure discussed in Section~\ref{s:size}. Here in order
to homogenize the system we apply a kind of a substitution
dynamics.

In Section~\ref{s:acceleration} we consider yet another
generalization of the well-known Nagel-Schreckenberg (NS)
traffic model \cite{NS} based on the introduction of a fractional
acceleration of particles made in \cite{Bl-hys}. Recall that the
original NS model already contained the acceleration term
which was assumed to be an integer. In \cite{Bl-hys} it has been
shown that the presence of the fractional acceleration leads
to very rich hysteresis type phenomena. Considered from a
bit more general point of view this model describes the
motion of particles on a lattice under the action of a
constant force represented by the acceleration. Therefore
if the particles in the configuration have different masses
or sizes it is natural to think that under the action of
the same force they will get different accelerations. This
is exactly the case we shall consider here. Again the
different accelerations might be the result of the reduction
of particle sizes.

\section{Dynamics of different size particles systems}\label{s:size}


Consider a class of locally interacting particle systems on the
one-dimensional integer lattice $\IZ$. Each particle is
described by its {\em position} -- the most left site $i$ which
it occupies on the lattice, its {\em length} $\ell$ describing
the number of lattice sites it occupies, its {\em velocity} $v$
and may be some other parameters. The dynamics is defined as
follows. At each site of the lattice there is an alarm-clock and
at time $t>0$ we consider only those particles which occupy
lattice sites where the alarm rings. For each such particle
with the position $i\in\IZ$ we calculate its new velocity
$v_i$ using a (random or deterministic) procedure which is the same
for all particles and does not depend neither on time nor on
the other particles in the configuration. About the velocity
we shall assume only that $|v_i|\le V_{\rm max}<\infty$. Here
$V_{\rm max}$ plays the role of the largest allowed velocity
and its boundedness defines the locality of interactions.

Then one checks a certain {\em admissibility} condition related
to the possibility to move a particle from the site $i$ to the
site $i+v_i$. We assume that the admissibility condition is again
local and depends only on the present positions of the particles
in a $2V_{\rm max}$ lattice neighborhood of the site $i$.
A natural assumption here is that the velocity $v_i$ is not
admissible if during the movement from the site $i$ to the site
$i+v_i$ the particle needs to go through an occupied site.
Only if the admissibility condition is satisfied the particle
is moved to a new position. Then for all sites to where the
particles were moved we restart the alarm-clocks (again using
a certain random or deterministic procedure).

We assume that the procedures used to choose new velocities
and to restart the alarm-clocks are the same for all sites
and do not depend on time.

Depending on the way how one restarts the alarm-clocks both
continuous and discrete time particle systems can be considered.
In what follows we restrict ourselves to a (more interesting
from our point of view and much less studied) discrete time
case, assuming that the alarm-clocks start with the same setting
and after each restart we add one to the time. Therefore all
particles are trying to move simultaneously.

A typical and well-known model satisfying our assumptions is the
so called exclusion process (see e.g. \cite{Lig}). A particular
deterministic case well suited for the description of traffic
flows will be studied in Section~\ref{s:acceleration}.

The scheme of the size reduction is as follows: we introduce a
very general dynamics of particles on an integer lattice where
the next position/velocity of a particle depends only on the
present position/velocity and on the positions of the particles
in its neighborhood. Then we construct a bijective map $\pi$ from
the space of original particle configurations to the space of
configurations of equal size particles which induces the new
dynamics. The idea is to map simultaneously each original
particle to a particle of size one and to delete all sites
occupied by the particle except for just one of them from the
lattice.
To make the formal definition of the application of the map $\pi$
to a configuration $x\in\X$ we enumerate all particles in this
configuration according to their natural order by integers and
setting the index $0$ to the particle occupying the smallest
nonnegative position on the lattice $\IZ$. Thus for each index
$j\in\IZ$ we know the position $i_j$ and the length $\ell_j$ of
the corresponding particle. The positions of unit size particles
in the configuration $\pi(x)$ we define recursively. First we
put a particle of unit size to the position $i_0$. After that we
choose the particle with the index $j=1$ in the original
configuration and put a particle of unit size to the position
$i_1-\ell_0+1$. The next particle to the right in the original
configuration corresponds to the index $j=2$ and will go to the
position $i_2-\ell_0+1-\ell_1$, etc. Thus the particle with the
index $j>0$ gets the position
$$ i_j' := i_j - \sum_{k=0}^{j-1}(\ell_k-1) .$$
Similarly one defines positions for the particles with negative
indices: if $j<0$ then
$$ i_j' := i_j + \sum_{k=j}^{-1}(\ell_k-1) .$$
Eventually we define the configuration $\pi(x)$ such that for
each $j\in\IZ$ there is a particle of unit size at site
$i_j'$ and all other sites are occupied by vacancies
(see Fig~\ref{f:recoding}).

\Bfig(300,50)
      {\bline(0,40)(1,0)(300)   \put(304,38){$x$}
       \put(10,40){\fbox{\qquad}} \put(60,40){\fbox{\qquad\qquad}}
       \bezier{30}(120,-2)(120,25)(120,45) \put(121,-10){$0$}
       \put(150,40){\fbox{\qquad\quad}} \put(215,40){\fbox{}}
       \put(250,40){\fbox{\qquad}}
       \bline(0,5)(1,0)(300)    \put(304,3){$\pi(x)$}
       \put(150,5){\fbox{}} \put(180,5){\fbox{}} \put(215,5){\fbox{}}
       \put(100,5){\fbox{}} \put(70,5){\fbox{}}
       \put(170,40){\vector(-1,-2){16}}
       \put(217,40){\vector(-1,-1){33}}
       \put(267,40){\vector(-3,-2){48}}
       \put(83,40){\vector(2,-3){22}}
       \put(26,40){\vector(4,-3){45}}
       \put(150,-10){$i_1'$} \put(180,-10){$i_2'$} \put(215,-10){$i_3'$}
       \put(104,-10){$i_{-1}'$} \put(73,-10){$i_{-2}'$}
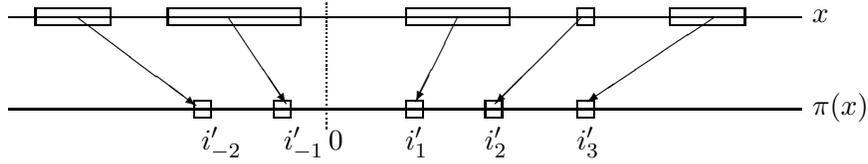
      }{Mapping of an inhomogeneous configuration to a homogeneous one.
        \label{f:recoding}}

Observe that the enumeration is preserved under the action of the
map $\pi$ and we assume that all parameters of the particles
except their sizes are preserved as well. Note also that the
enumeration makes the map $\pi$ invertible, so this is a
bijection. Nevertheless the total number of sites in a finite
segment of the lattice may become smaller after the action of the
map $\pi$.

Since in the model under consideration the dynamics does not depend
on the sizes of the particles, the dynamics of the system after
the application of the map $\pi$ is preserved. Therefore if we
know any statistical description of the new homogeneous particle
system (e.g., its invariant measure, correlation functions, etc.)
then applying the inverse map $\pi^{-1}$ we get immediately the
corresponding description of the original system.

Despite giving the complete information about the conjugation
between the original inhomogeneous model and the resulting
homogeneous one the map $\pi$ is very complicated and it is
desirable to have a more direct way to derive relations between
various statistics of the original and the resulting systems. As
we shall show at least for some important statistics, such as
limit average particle velocities, the presence of the bijection
gives a possibility to calculate them rigorously using only the
average size of particles and their density.

Denote the space of positions corresponding to particles in
admissible particle configurations by $\X\subset\{0,1\}^{\IZ}$.
There is an important property that holds for all systems we
consider here: particle conservation. For a configuration
$x\in\X:=\{0,1\}^{\IZ}$ and a finite subset $I\subset\IZ$ denote
by $\den(x,I)$ the number of particles from the configuration $x$
located in $I$ divided by the total number of sites in $I$.
Clearly $0\le\den(x,I)\le1$. Choosing a sequence of lattice
segments $I_n$ of length $n$ we consider the limit
$\lim_{n\to\infty}\den(x,I_n)$. If this limit exists and does not
depend on the sequence $\{I_n\}_n$ we call it the {\em particle
density} of the configuration $x\in\X$ and denote by $\den(x)$. To
show that the particle density is conserved under dynamics
consider a segment of sites $I$ of length $L$ and denote by $N_p$
and $N_p'$ the numbers of particles in this segment at time $t$
and by $t+1$. During one time step at most two particles may leave
or enter the segment $I$ and thus $|N_p - N_p'|\le2$. Therefore
passing to the limit as $L\to\infty$ we get the conservation of
the density.

To this end let us introduce the notion of the {\em average
velocity} of a particle. Let $L(x,i,t)$ be the distance covered
during the time $t$ by a particle in the configuration $x$ located
initially (at $t=0$) at the site $i\in\IZ$. Then by the average
velocity of this particle we mean
$$ V(x,i,t) := \frac1t L(x,i,t) $$
and consider also the limit average velocity
$$ V(x,i):=\lim_{t\to\infty} V(x,i,t) $$
provided that it is well defined, otherwise one considers limit
points of the sequence $\{V(x,i,t)\}_t$. In
Section~\ref{s:acceleration} we shall show that under some natural
assumptions the above limit is well defined and is the same for
all particles. In general this might not be the case, nevertheless
we shall show that all limit points of the average velocities in
the original system can be easily calculated from the
corresponding limit points obtained for the homogeneous model.

To simplify notation we shall use the sign ``prime'' for various
parameters of particles related to the configuration $x':=\pi(x)$.
Choose a particle in the configuration $x$ having a position
$i\in\IZ$ and compare the distance $L(x,i,t)$ which it covers
during the time $t>0$ to the distance $L'(x',i',t)$ covered by the
corresponding particle located initially at $i'\in\IZ$ in the
homogeneous model. For a positive integer $\ell'$ denote by
$N_p(\ell')$ and $N_v(\ell')$ the number of particles and
vacancies respectively in the segment $[i',i'+\ell'-1]$ of th
configuration $x'$. Set
$$ s(\ell'):=(\ell'-N_v(\ell'))/N_p(\ell') $$
and assume that the limits
$$ s:=\lim_{\ell'\to\infty}s(\ell'), \qquad
   \den:=1 - \lim_{\ell'\to\infty} N_v(\ell')/\ell, \qquad
   \den':=\lim_{\ell'\to\infty}N_p(\ell')/\ell' $$
do exist. The first of them is the average size of a particle in
the original configuration and the other two are the densities of
particles in the original and the homogeneous models. Therefore
the limits certainly exist if the configuration $x'$ has the
particle density. Note that in the definition of the value $\den$
the denominator is equal to $\ell$ (corresponding to the
inhomogeneous system) rather than $\ell'$.

Let $V'(x',i')$ be a limit point of the average velocities in the
homogeneous system, i.e. there exists a sequence of moments of
time $t_k\toas{k\to\infty}\infty$ such that
$$ V'(x',i'):=\lim_{k\to\infty}L'(x',i',t_k)/t_k .$$
For any $t>0$ one has
$$ L(x,i,t) = s(L'(x',i',t))\cdot N_p(L'(x',i',t)) + N_v(L'(x',i',t)) $$
while
$$ L'(x',i',t) = N_p(x',i',t) + N_v(x',i',t) .$$
Thus
$$ L(x,i,t) - L'(x',i',t) = (s(L'(x',i',t))-1)\cdot N_p(L'(x',i',t) .$$
Therefore %
$$ \frac{L(x,i,t_k)}{t_k} = \frac{L'(x',i',t_k)}{t_k} \cdot
                          \frac{L(x,i,t_k)}{L'(x',i',t_k)}
   \toas{k\to\infty} (1+(s-1)\den')\cdot V'(x',i') .$$ %
Hence passing the average velocities to the limit along the same
sequence of moments of time $\{t_k\}$ we get%
\beq{e:v-v}{ V(x,i) = (1+(s-1)\den')\cdot V'(x',i') .} %
Similarly, but much simpler one gets the relation between
the particle densities in the original and the corresponding homogeneous
systems: %
\beq{e:r-r}{\den:= 1-\lim_{\ell\to\infty}\frac{N_v(\ell')}\ell
      = 1 - (1-\den')\cdot\lim_{\ell\to\infty}\frac{\ell'}\ell
      = \frac{s\den'}{1+(s-1)\den'} .}%
Note that both these results do not depend neither on the fine
features of the dynamics under consideration nor on the details of
the distribution of particle sizes. Observe also that a naive idea
to construct the relation between the average velocity and the
particle density in the original system in terms of the density of
the occupied sites using the corresponding formulae known for the
homogeneous system does not work.

\section{Different maximal velocities}\label{s:max-vel}

In the model we have discussed in the previous section there was
a parameter $V_{\rm max}$ describing the maximal available
velocity of particles. Assume now that this parameter varies from
one particle to another but each velocity may take only two
values: $0$ or the corresponding positive maximal value. Our aim
is to show that one can construct a new bijection $\cC$ between
the inhomogeneous system with particles having different
velocities and a homogeneous one having only identical `slow'
particles having the unit maximal velocities.

This can be done as follows: %
$$  ... 0 0 0 1_2 0 0 0 0 0 0 1_1 0 0 1_4 0 0 0 0 0 0 0 ... ~~\to~~ %
   ... 0_3 1_2 0_2 0_4  1_1 0_1 0_1 1_4 0_4 0_3 ... $$ %
Here in the left representation $0$ stands for a vacancy and $1$ with
and the index $v$ denotes a particle with the maximal available velocity
$v$. Under the dynamics a particle ``exchanges'' its position with
a certain number of succeeding vacancies. Therefore in the second
representation we code both particles and vacancies by groups represented
by $0_v$ and $1_v$. The indices corresponding to vacancies play here
a different role: a zero with an index $v$ means $v$ vacancies in the
original configuration.
The index of $0$ is the minimum between the index the preceding
$1$ and the total number of vacancies in the original
configuration immediately after the particle, i.e. %
$$ ...1_20001_301_10000... \to ...1_2 0_2 1_3 0_1 1_1 0_1 0_3 ...$$ %
The dynamics of the new system again consists of exchanges of
particles and vacancies.

To make the above coding precise we introduce the alphabet
$\cA:=\{0_1,0_2,\dots,0_{V_{{\rm max}}},
        1_0,1_1,\dots,1_{V_{{\rm max}}}\}$ with $2V_{{\rm max}}+1$
elements and a map $\cC:X\to\cA^{\IZ}$ defined through the
following system of substitutions: %
$$ 1_v\underbrace{0\dots0}_{n}1_{v'} ~\toas{\cC}~
   1_v\underbrace{0_v\dots0_v}_{\intp{n/v}}0_{n-\intp{n/v}v}1_{v'} \CR%
   \underbrace{0\dots0}_{\infty}1_{v} ~\toas{\cC}~
   \underbrace{0_{v}\dots0_{v}}_{\infty}1_{v}  \qquad %
   1_v\underbrace{0\dots0}_{\infty} ~\toas{\cC}~
   1_v\underbrace{0_{v}\dots0_{v}}_{\infty}
   .$$
Here $v$ is the maximal velocity of the particle immediately
preceding the block of zeros and $\intp{\cdot}$ stands for the
integer part of a number. The last two relations define the action
of $C$ on `tails' of $x$ consisting entirely of vacancies. In
other words a configuration is divided into blocks of consecutive
vacancies surrounded by particles and each of the blocks is
substituted by a block of zeroes indexed by the maximal velocity
of the particle immediately preceding this block (except for the
last indexed zero where the index is calculated as the remainder)
according to the above substitution rules.

As we see the dynamics of the original system is equivalent
through the bijection $\cC$ to the dynamics of the composition of
the dynamics with `slow' particles composed with $\cC$. Using the
approach similar to the one developed in the previous Section one
can calculate explicit relations between the limit points of
average velocities in the original system and in the constructed
homogeneous one.

Note that a similar idea of the substitution dynamics has been
applied earlier in another author's paper \cite{Bl-erg} to reduce
the analysis of a deterministic homogeneous model of a traffic
flow with `fast' particles to the `slow' particles case.

\section{A traffic model with different particle accelerations}
\label{s:acceleration}
Let $x$ be a configuration of particles on the one-dimensional
integer lattice $\IZ$ having at most one particle at a site.
To each particle we associate two real variables:
$0\le v\le V_{\rm max}<\infty$ (which we call {\em velocity}) and
$0\le a\le1$ (which we call {\em acceleration}).
A configuration is called {\em admissible} if each particle
can be moved by the distance equal to the integer part of
its velocity (notation $\intp{v}$) not interacting with
other particles in the configuration.

The dynamics is defined as follows. First we modify the
velocities adding to each of them the corresponding acceleration
and observing the restriction that velocities cannot exceed
$V_{\rm max}$. In order to satisfy the admissibility condition we
compare each of the resulting velocities to the distance to the
next particle to the right (denote it by $\ell$) and take the
minimum if needed. Thus the modified velocity can be written as
follows: $\min\{v+a,V_{\rm max},\ell\}$. After that each particle
is moved to the right by the distance equal to the integer part
of its velocity $\intp{v}$ (see Fig.~\ref{f:free-motion}). It is
immediate to check that this model satisfies all properties
assumed in Section~\ref{s:size}.

\Bfig(150,50)
      {
       \bline(0,40)(1,0)(150)   \put(154,38){$t$}
       \put(10,40){\circle*{5}} \put(10,48){\vector(1,0){40}}
       \put(45,40){\circle*{5}} \put(45,45){\vector(1,0){47}}
       \bline(0,0)(1,0)(150)    \put(154,-2){$t+1$}
       \put(35,0){\circle*{5}} 
       \put(85,0){\circle*{5}} \put(85,5){\vector(1,0){47}}
       \bezier{30}(35,0)(35,22)(35,45)
       \bezier{30}(85,0)(85,22)(85,45)
       \put(8,28){$i$} \put(43,28){$j$}
       \put(10,50){$v_{i}+a_{i}$}  \put(55,50){$v_{j}+a_{j}$}
       \put(20,-12){$i'=j-1$} \put(83,-12){$j'$}
       \put(38,10){$v'_{i}=0$}  \put(88,10){$v'_{j}=v_{j}+a_{j}$}
      }{Dynamics of particles acceleration: $i<j$ and
        $i'<j'$ -- the positions of neighboring particles and
        $v_i, v_j, v'_{i},v'_{j}$ -- the corresponding velocities
        at time $t$ and $t+1$.
        \label{f:free-motion}}

Under the restriction $V_{\rm max}=1$ and all accelerations are
identical ergodic properties of this model have been studied in
\cite{Bl-hys}.\footnote{In \cite{Bl-hys} the movement of particles
    and their acceleration were applied in the opposite order but
    this makes no difference for statistical properties of the
    system.} %
In the limiting case when the acceleration is equal to one this
model coincides exactly with the well-known NS model, while for
fractional values of the acceleration it imitates some more
complicated non Markov traffic models (see a discussion in
\cite{Bl-hys}). In \cite{Bl-hys} it has been shown that that the
fractional acceleration leads to a very rich hysteresis type
phenomena.

Considered from a bit more general point of view this model
describes the motion of particles on a lattice under the action
of a constant force represented by the acceleration $a$, while
the presence of the finite maximal velocity may be interpreted as
a result of viscosity. Therefore if the particles in the
configuration have different masses or sizes it is natural to
think that under the action of the same force they will get
different accelerations. This is exactly the case we consider
here. Fig.~\ref{av-speed-den} summarizes the results of
\cite{Bl-hys} describing the dependence between the average
velocity and the particle density of a homogeneous system of
particles with $V_{\rm max}=1$ and $a\le1$. Here
$\gamma_1:=(1+\intpm{1/a}V_{\rm max})^{-1}$, $\gamma_2:=(1+V_{\rm
max})^{-1}$, and $\intpm{\cdot}$ stands for the smallest integer
not smaller than the considered number.

\Bfig(100,100)
      {\footnotesize{
       \bline(0,0)(1,0)(160)   \bline(0,0)(0,1)(100)
       \bline(0,100)(1,0)(160) \bline(160,0)(0,1)(100)
       \put(20,20){\vector(1,0){120}} \put(20,20){\vector(0,1){70}}
       \thicklines
       \bline(20,75)(1,0)(45)  \bezier{200}(35,75)(70,23)(125,20)
       \thinlines
       \bezier{25}(65,75)(65,47)(65,20)
       \put(145,18){$\rho$}    \put(32,10){$\gamma_{1}$}
       \put(123,12){1}
       \put(12,85){$V$}        \put(0,72){$V_{\rm max}$}
       \put(80,37){$C(\frac1\rho-1)$}
       \put(63,10){$\gamma_{2}$}
       \bezier{30}(35,75)(35,50)(35,20)
      }}
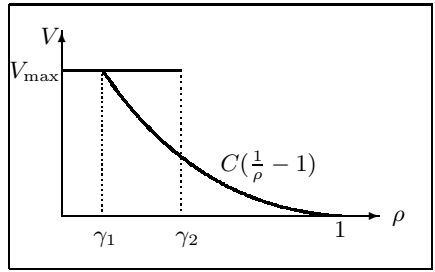
{Dependence of the limit average velocity $V$ on the density of
particles $\rho$. \label{av-speed-den}}

The most interesting part of Fig.~\ref{av-speed-den} corresponds
to the region of densities between two critical values
$\gamma_{i}$ where the one to one correspondence between the
average velocity and the density breaks down. In fact, the
correspondence in this region is even more complicated and we
refer the reader for the detailed analysis to \cite{Bl-hys}.


Clearly in the absence of obstacles the particles in the
configurations are moving freely under their accelerations until
they get the largest velocity. All peculiarities of the traffic
are connected to `jams' (when the motion of a particle is blocked
by another one due to the admissibility condition) as the only
possible obstacles to the free motion of particles.

We shall say that a {\em jam} $J$ is a locally maximal collection
of consecutive particles in a given configuration having velocities
strictly smaller than the maximal allowed one $V_{\rm max}$.

The number of particles and their positions in a jam may change
with time: leading particles are becoming free (i.e. getting the
maximal allowed velocity $V_{\rm max}$) and some new particles are
joining the jam coming from behind. However, only
one such change at a time might happen, and, in particular, a jam
cannot split into several new jams. Therefore we can analyze how
a given jam changes with time and the main quantity of interest
for us here is the minimal number of iterations after which the
jam will cease to exist. Denote by $J(t)$ the segment
corresponding to the given jam at the moment $t$ (in this
notation $J(0)$ is the original jam). Then by the {\em life-time}
of the jam $J$ we shall mean %
\beq{e:life-def}{\tau(J):=\sup\{t: ~ |J(t)|>0, ~ t>0\} ,}%
where $|A|$ is the length of the segment $A$.

`Attracting' the preceding particles, a jam plays a role similar
to an attractor in dynamical systems theory. Therefore it is
reasonable to study it in a similar way and to introduce the
notion of the {\em basin of attraction} (notation BA$(J)$) of the
jam $J$, by which we mean the minimal segment of the configuration
$x$ containing all sites from where particles may eventually join
the jam.

The example on Fig~\ref{ex:model-dyn} demonstrates the dynamics
of a jam consisting initially of 3 particles with zero velocities
in a system with $V_{\rm max}=2$ and identical accelerations
$a=1/2$. We indicate the positions of particles by their
velocities and $b$ stands for the velocity $3/2$. Dots indicate
the positions of vacancies belonging to the BA of the jam, and $t$
corresponds to time. In each line all marked positions up to the
last particle having velocity strictly less than $2$ belongs to
the basin of attraction of the jam. The example shows that the
left boundary of a BA moves at constant velocity $V_{\rm max}$
which follows immediately from its definition, but its right
boundary coinciding with the leading particle of the jam
fluctuates quite irregularly even in this simple example.

\begin{figure} \begin{center}
\begin{tabular}{lllllllllllllllllllllll}
t=0&.&.&.&.&.&.&.&.&.&.&.&.&.&0&.&0&0& & & & & \\
t=1& & &.&.&.&.&.&.&.&.&.&.&.&a&.&0&a& & & & & \\
t=2& & & & &.&.&.&.&.&.&.&.&.&.&1&0&.&1& & & & \\
t=3& & & & & & &.&.&.&.&.&.&.&.&0&a&.&.&b& & & \\
t=4& & & & & & & & &.&.&.&.&.&.&0&.&1& & & &2& \\
t=5& & & & & & & & & & &.&.&.&.&a&.&.&b& & & & \\
t=6& & & & & & & & & & & & &.&.&.&1& & & &2& & \\
t=7& & & & & & & & & & & & & & &.&.&b& & & & & \\
t=8& & & & & & & & & & & & & & & & & & &2& & & \\
\end{tabular}
\end{center}%
\vskip-0.5cm \caption{An example of the dynamics with $V_{\rm
max}=2, a=\frac12, b=\frac32$. The positions of particles are
marked by their velocities and the positions of vacancies
belonging to the BA by dots. \label{ex:model-dyn}}
\end{figure}%

It has been shown in \cite{Bl-hys} that in the case of the
constant acceleration and $V_{\rm max}=1$ the knowledge of the
positions of particles in the BA(J) gives the exact value of
$\tau(J)$. It is important that in that case only {\em static}
jams in which all particles (except the leading one) have zero
velocities are possible. If $V_{\rm max}>1$ there might be {\em
dynamic} jams where all particles are moving at velocities
strictly less than $V_{\rm max}$ which makes the calculation of
the life-time much more complex. Moreover the life-time in this
case depends not only on the positions of the particles but on
their velocities as well. Of course, varying accelerations make
the situation even more complicated.

Nevertheless we shall show that even without the information
about the velocities one can get the upper estimate for the
life-time of a cluster which will be sufficient for us.

Let at time $t_0$ the BA of a jam $J$ under consideration
consists of $m$ consecutive particles indexed according
to their positions by numbers $i\in\{1,\dots,m\}$ and
denote the corresponding accelerations by $a_i$. Then the
leading particle having the index $m$ gets the maximal
velocity $V_{\rm max}$ at most after
$\intpm{V_{\rm max}/{a_m}}$ time steps.\footnote{
   Actually the number of time steps is equal to
   $\intpm{(V_{\rm max}-v_m)/{a_m}}$ where $v_m$ means
   the initial velocity of the leading particle.}
After at most another $\intpm{V_{\rm max}/{a_{m-1}}}$ time
steps the second leading particle gets the maximal velocity
etc. Therefore we get the upper estimate for the life-time:
$$ \tau(J) \le \sum_{i=1}^m \intpm{V_{\rm max}/{a_i}} .$$
In fact, this estimate is optimal if $V_{\rm max}=1$.
If $V_{\rm max}>1$ several particle may accelerate
simultaneously which might diminish the life-time
significantly.

Denoting $a(x):=\inf a_i$, where the infimum is taken over
accelerations of all particles present in the configuration $x$,
and assuming that $a(x)>0$ we get that the life-time
of a jam $J$ in this configuration satisfies the inequality %
\beq{e:life-time}{\tau(J)
  \le \intpm{V_{\rm max}/a(x)} \cdot |BA(J)| .}%
Here $|BA(J)|$ stands for the number of particles in
the basin of attraction of the jam $J$.

Consider now what is happening on the level of individual
particles in a configuration having no infinite life-time jams.
Clearly each particle can move through a jam spending there only
a finite time, but it might be possible that ahead of a given
particle there is an infinite sequence of jams with monotonously
growing (albeit finite) life-times. Therefore some additional
care is needed to show that this does not prevent the particle to
become eventually free (i.e. moving at maximal velocity).

Recall that the average velocity of a particle in a configuration
$x$ located initially (at $t=0$) at the site $i\in\IZ$ is defined
as
$$ V(x,i) :=\lim_{t\to\infty}\frac1t L(x,i,t) $$
provided that it is well defined. Here $L(x,i,t)$ is the distance
covered by this particle during the time $t$. A simple argument
similar to Lemma~2.2 in \cite{Bl-hys} shows that if the above
limit is well defined for a certain particle in the given
configuration then it does not depend on the initial coordinate
of the particle and is the same for all particles. Choose any
pair of consecutive particles located initially at sites $i<j$.
Under dynamics the distance between these particles changes
according to the difference between their velocities, which might
take values between $0$ and $V_{\rm max}$. Since the left
particle can be slowed down only by the right one, we see that
for any moment of time $t$ the distance between the particles can
be enlarged at most by $CV_{\rm max}$, where the constant
$C<\infty$ depends on the accelerations
of these particles but not on time. Thus %
$$ 0 \le (j + L(x,j,t)) - (i + L(x,i,t))
     \le j - i + CV_{\rm max} $$
or
$$ j - i \le L(x,j,t) - L(x,i,t) \le CV_{\rm max} .$$
Dividing by $t$ and using the definition of the time average
velocity we get %
\beq{eq:vel-inc}{
 |V(x,i,t) - V(x,j,t)|
 \le \max\left\{\frac{CV_{\rm max}}t, ~~ \frac{j - i}t\right\}
 \toas{t\to\infty}0 .}%
Thus $V(x,i)=V(x,j)$. Using the same argument one extends this
result to neighboring particles, and repeating it to all
particles in the configuration. Therefore we may drop the
index $i$ in the definition of the average velocity.

In order to show that the absence of infinite life-time jams leads
to the free eventual motion of particles it is enough to prove
that $V(x)=V_{\rm max}$. Consider a partition of the integer
lattice by nonintersecting finite BAs corresponding to jams in the
configuration $x$ and their complement (gaps between the jams).
Choose one of those BAs and denote by $i$ the position of the
first particle preceding it. According to the definition of the BA
this particle will never join the jam corresponding to the BA.
Moreover, since we assume that the BAs included in the partition
are disjoint, the particle never join any jam corresponding to the
elements of the partition. Thus it will have no obstacles in its
motion and after at most $\intpm{V_{\rm max}/a}$ time steps it
will get the maximal velocity $V_{\rm max}$.

This together with the independence of the average velocity
on the initial position of the particle proves that
$V(x)=V_{\rm max}$.

It remains to discuss the connection between the condition
of the absence of infinite life-time jams and the particle
density.

In principle, arguments used in the proof of Lemma~3.1
of \cite{Bl-hys} can be extended to the case of `variable'
accelerations but only under the condition $V_{\rm max}=1$.
Even in this case the calculations are becoming rather
messy. Therefore instead of getting more sharp estimates
working only in the case of `slow' particles we shall
obtain much more rough estimates of the critical
density under which there are no infinite life-time jams
being valid for any $V_{\rm max}$.

Consider an infinite life-time jam $J=x[m,n]$. Recall that
according to the definition of a jam a free particle located
initially at the site preceding the BA of a jam cannot join the
jam. Therefore using the estimate for the life-time of a jam
(\ref{e:life-time}) one can show that if the segment $x[n-L+1,n]$
is contained in the BA of this jam then
$$ N_v < N_p \intpm{V_{\rm max}/a(x)} .$$
Here $N_p$ is the number of particles in the segment
$x[n-L+1,n]$ and $N_v:=L-N_p$ is the number of vacancies
in the segment. Therefore
$$ \frac{N_p}L =  \frac{N_p}{N_p+N_v}
 > (\intpm{V_{\rm max}/a(x)}+1)^{-1} .$$
Passing to the limit as $L\to\infty$ and using that the
BA of this jam is infinite as well we come to the estimate
of the critical density below which there are no infinite
life-time jams: %
\beq{e:crit-den}{\den(x) \ge
                 \gamma_1:=\frac1{\intpm{V_{\rm max}/a(x)} +1} .}%

Interestingly, the upper bound $\gamma_2$ of the densities when
there exist configurations consisting of only free moving
particles does not depend on the accelerations. To calculate
$\gamma_2$ consider the most `compressed' configuration of free
particles. Since each of them is moving at velocity $V_{\rm max}$
then the distance between neighboring particles cannot be smaller
than $V_{\rm max}$, which immediately yields $\gamma_2:=(V_{\rm
max}+1)^{-1}$.

Similarly to the results of \cite{Bl-hys} corresponding to the
case of the constant acceleration and `slow' particles with
$V_{\rm max}=1$ one expects that the model under considerations
has two distinct ergodic (unmixed) phases with two critical
values of the particle density. When the density is below the
lowest critical value, the steady state of the model corresponds
to the ``free-flowing'' (or ``gaseous'') phase. When the density
exceeds the second critical value the model produces large,
persistent, well-defined traffic jams, which correspond to the
``jammed'' (or ``liquid'') phase. Between the two critical values
each of these phases may take place, which can be interpreted as
an ``overcooled gas'' phase when a small perturbation can change
drastically gas into liquid.

The estimates we obtained so far correspond to the ``gaseous''
phase. It can be shown that when the particle density exceeds the
second critical value $\gamma_2$ not only the jams are unavoidable
but infinitely many infinite life-time jams are present. This
explains why we call this phase as ``jammed''. As we already
mentioned high maximal velocity $V_{\rm max}>1$ leads to the
appearance of dynamic jams completely absent in the case of `slow'
particles, which in turn complicates the analysis of the region
between the critical values. Clearly the presence of different
accelerations leads to even more complicated dependence between
the average velocities and the particle densities and as we expect
without the knowledge of the distribution of the accelerations one
cannot derive this dependence.

Recalling the dependence of the average velocity and the density
in the ``liquid'' phase obtained exactly in the homogeneous case
(see Fig~\ref{av-speed-den}) we see that it depends heavily on
the acceleration. Therefore in the non homogeneous case one cannot
expect to get any functional dependence here.

\section{Conclusion}
In this paper it has been shown that in a number of cases the
analysis of non homogeneous lattice interacting particle systems
(IPS) may be reduced to the analysis of homogeneous ones. In
distinction to known examples of this sort where the inhomogeneity
was due to varying hopping rates (so one expects a certain self
averaging) we have shown that a rigorous reduction may be achieved
even when the sizes of particles are different. Interestingly, in
the latter case the details of the size distribution does not play
an important role (see the derivation of the limit average
velocity and density in Section~\ref{s:size}). Additionally we
have considered two different situation when the inhomogeneity
come in the form of varying particle velocities or accelerations
(which can be considered as a version of varying hopping rates).
In both cases we obtained either an exact reduction to the
homogeneous system or rigorous estimates of important statistical
quantities.

The analysis made in the paper was restricted to the discrete
time systems, i.e. to IPS with the parallel updating. On the
other hand, all results of Sections~\ref{s:size} and
\ref{s:max-vel} hold in complete generality both for continuous
time IPS and systems with random sequential updating.

To finalize let us describe a few open problems in the field.

Due to the clear connections of the IPS under study to traffic
flow modelling it would be of interest to extend our results about
the mapping of the IPS with particles of different size to
multilane traffic models, where the particles are moving and
exchanging positions along several lattice lines. Multilane
traffic models with identical particles were studied
mathematically, e.g. in \cite{Bl-erg}, where the exact dependence
between limit average velocities and particle densities were
obtained. We expect that a version of the size reduction developed
in Section~\ref{s:size} should work here but the problem with the
non homogeneous case is that it is not clear how to take into
account the change of particles of different sizes between the
lanes.

Another set of questions is related to random versions of
the deterministic IPS discussed in Section~\ref{s:acceleration},
when the movement of particles happen with a certain (may be
non homogeneous again) probability. At the moment nothing is
known rigorously about such systems and it would be of interest
to prove the existence and uniqueness of invariant distributions
corresponding to each value of the particle density in the true
probabilistic setting.

Throughout the paper we have considered only particle
configurations having densities, i.e. being spatially ergodic
with respect to the standard shift-map. Applying the ideas
developed in \cite{Bl-hys, Bl-erg} one can extend our results to
a more general setting using lower and upper densities instead of
the usual density. From this point of view it is of interest to
study limit statistics corresponding to particle configurations
having different {\em left} and {\em right} densities: $\rho_{\rm
left}(x):=\lim_{n\to\infty}\frac1n N_p(x[-n,-1])=\alpha$,
$\rho_{\rm right}(x):=\lim_{n\to\infty}\frac1n N_p(x[1,n])=\beta$.
One can think about this as an imitation of `open' systems with
the entrance rate $\alpha$ and the exit rate $\beta$.

\section*{Acknowledgments}
This research has been partially supported by Russian Foundation
for Fundamental Research, CRDF and French Ministry of Education
grants. The author would like to thank Rahul Pandit and an
anonymous referee for very useful comments.


\end{document}